\documentclass[letterpaper,english,reprint, aps]{revtex4-1}
\usepackage{Riemann_function}
\usepackage{graphicx}
\usepackage{epsfig}
\usepackage{amsmath,amsfonts,amsthm}
\usepackage{amssymb}
\usepackage{times}

\newcounter{defcounter} 
\newenvironment{myequation}{%
\addtocounter{equation}{-1} 
\refstepcounter{defcounter} 
 
\begin{equation}} 
{\end{equation}}

\usepackage[T1]{fontenc}
\usepackage[latin9]{inputenc}
\usepackage{color}
\usepackage{babel}
\usepackage{amsmath}
\usepackage{amssymb}
\usepackage{graphicx}
\usepackage{mathtools}
\usepackage[unicode=true,pdfusetitle,
 bookmarks=true,bookmarksnumbered=false,bookmarksopen=false,
 breaklinks=false,pdfborder={0 0 1},backref=false,colorlinks=true]
 {hyperref}
 \usepackage{ragged2e}
\usepackage{caption}

\pdfpageheight\paperheight
\pdfpagewidth\paperwidth

\begin{document}

\title{Identifying the Riemann zeros by periodically driving a single qubit}

\author{Ran He$^{1,2}$}

\author{Ming-Zhong Ai$^{1,2}$}

\author{Jin-Ming Cui$^{1,2}$}
\email{jmcui@ustc.edu.cn}

\author{Yun-Feng Huang$^{1,2}$}
\email{hyf@ustc.edu.cn}

\author{Yong-Jian Han$^{1,2}$}
\email{smhan@ustc.edu.cn}

\author{Chuan-Feng Li$^{1,2}$}
\email{cfli@ustc.edu.cn}

\author{Tao Tu$^{1,2}$}

\author{C.E. Creffield$^{3}$}

\author{G. Sierra$^{4}$}

\author{Guang-Can Guo$^{1,2}$}

\affiliation{$^{1}$CAS Key Laboratory of Quantum Information, University of Science
and Technology of China, Hefei, 230026, People's Republic of China.}
\affiliation{$^{2}$CAS Center For Excellence in Quantum Information and Quantum Physics, University of Science and Technology of China, Hefei, 230026,
People's Republic of China.}
\affiliation{$^{3}$Departamento de F\'isica de Materiales, Universidad Complutense de Madrid, E-28040, Madrid, Spain.}
\affiliation{$^{4}$Instituto de F\'isica Te\'orica, UAM-CSIC, E-28049, Madrid, Spain.}

\begin{abstract}
\textbf{The Riemann hypothesis, one of the most important open problems in pure mathematics, implies the most profound secret of prime numbers. One of the most interesting approaches to solve this hypothesis is to connect the problem with the spectrum of the physical Hamiltonian of a quantum system. However, none of the proposed quantum Hamiltonians have been experimentally feasible.Here, we report the first experiment to identify the first non-trivial zeros of the Riemann zeta function and the first two zeros of P\'olya's fake zeta function, using a novel Floquet method, through properly designed periodically driving functions. According to this method, the zeros of these functions are characterized by the occurrence of crossings of quasi-energies when the dynamics of the system are frozen. The experimentally obtained zeros are in excellent agreement with their exact values. Our study provides the first experimental realization of the Riemann zeros, which may provide new insights into this fundamental mathematical problem.}{\normalsize\par}
\end{abstract}


\maketitle


\bigskip
\noindent\textbf{INTRODUCTION}

The Riemann hypothesis (RH) \citep{riemann1859ueber} states that all non-trivial zeros
of the Riemann zeta function $\zeta(s)$ are on the critical line of $Re[s]=1/2$ in the complex plane, i.e., all non-trivial zeros have
the form $s_{n}=1/2+iE_{n}$, where $E_{n}$ are real numbers. The RH is one of the most important unsolved problems
in the field of pure mathematics because it is closely connected to many fundamental
mathematical problems \citep{Siegel1932Uber,montgomery1973pair} such as the distribution of prime numbers\citep{Hardy1916Contributions}.

At a numerical level, more than $10^{13}$ zeros of the Riemann zeta function have been indentified using classical computer\citep{gourdon20041013}. However, a complete proof of the RH still needs to explored. At an experimental level, several physical implementations have been performed to simulate the zeros, such as an electronic analog device \citep{vanderpol1947}, and proposed in both classical systems, such as by the angular nodes separating the side-lobes of a far-field diffraction pattern, e.g. in optics or acoustics \citep{Berry2012Riemann,Berry2015Riemann}, and quantum systems, such as by two entangled quantum systems \citep{Feiler2013Entanglement}. Among all approaches to sovle the RH, the P\'olya--Hilbert conjecture (PHC) is a fascinating one, which supposes the existence of a Hamiltonian whose spectrum is the imaginary part of the Riemann zeros, and therefore real numbers. In this case all the Riemann zeros will be of the form $1/2 + iE_{n}$  with $E_{n}$ real, which is the statement of the Riemann hypothesis (RH). This conjecture was first proposed by P\'olya and Hilbert who were inspired
by the physical basis of the RH and the fact that all eigenvalues
of a physical Hamiltonian are real \citep{Polya1926Bemerkung, Edwards1974Riemann}. Because of the unexpected association between
the RH and quantum physics, and particularly the supporting evidence
from random matrix theory \citep{mehta2004random,firk2009nuclei,crisanti2012products}
and quantum chaos \citep{berry1990rule,bogomolny2007riemann}, several researchers have attempted to construct a suitable quantum Hamiltonian \citep{berry1999riemann,kuipers2014quantum,schumayer2008quantum,berry2011compact,schumayer2011colloquium,torosov2013quantum}. 
The $xp$ model is a well-known example \citep{berry1999riemann,sierra2011h,sierra2016riemann}, however, to our knowledge, no such Hamiltonian has been implemented in a real quantum system, e.g., more recently an operator related to the $xp$ Hamiltonian  whose eigenvalues correspond exactly to the Riemann zeros has been found \citep{bender2017hamiltonian}, but unfortunately it is not hermitian and therefore it is not a properly a Hamiltonian. 

Recently, a very different approach along these lines was proposed by Creffield and Sierra \citep{creffield2015finding}. In fact, a correspondence was established between the non-trivial zeros of the Riemann $\varXi$
function and the degeneracy of the quasi-energies in a periodically driven
qubit (rather than the eigenvalues of a static quantum Hamiltonian). This approach extends
the PHC to a driven system and sheds new light upon the association between
the RH and quantum physics. In particular, the periodic-driving
approach is experimentally achievable, although it requires demanding experimental conditions such as a long coherence time, low operational error, and low state-preparation
and measurement (SPAM) error. In this study, we report an experiment to precisely identify the first non-trivial zero of the Riemann zeta function and the first two zeros of P\'olya's function utilizing this Floquet approach.

\begin{figure}[t]

\includegraphics[width=1\columnwidth]{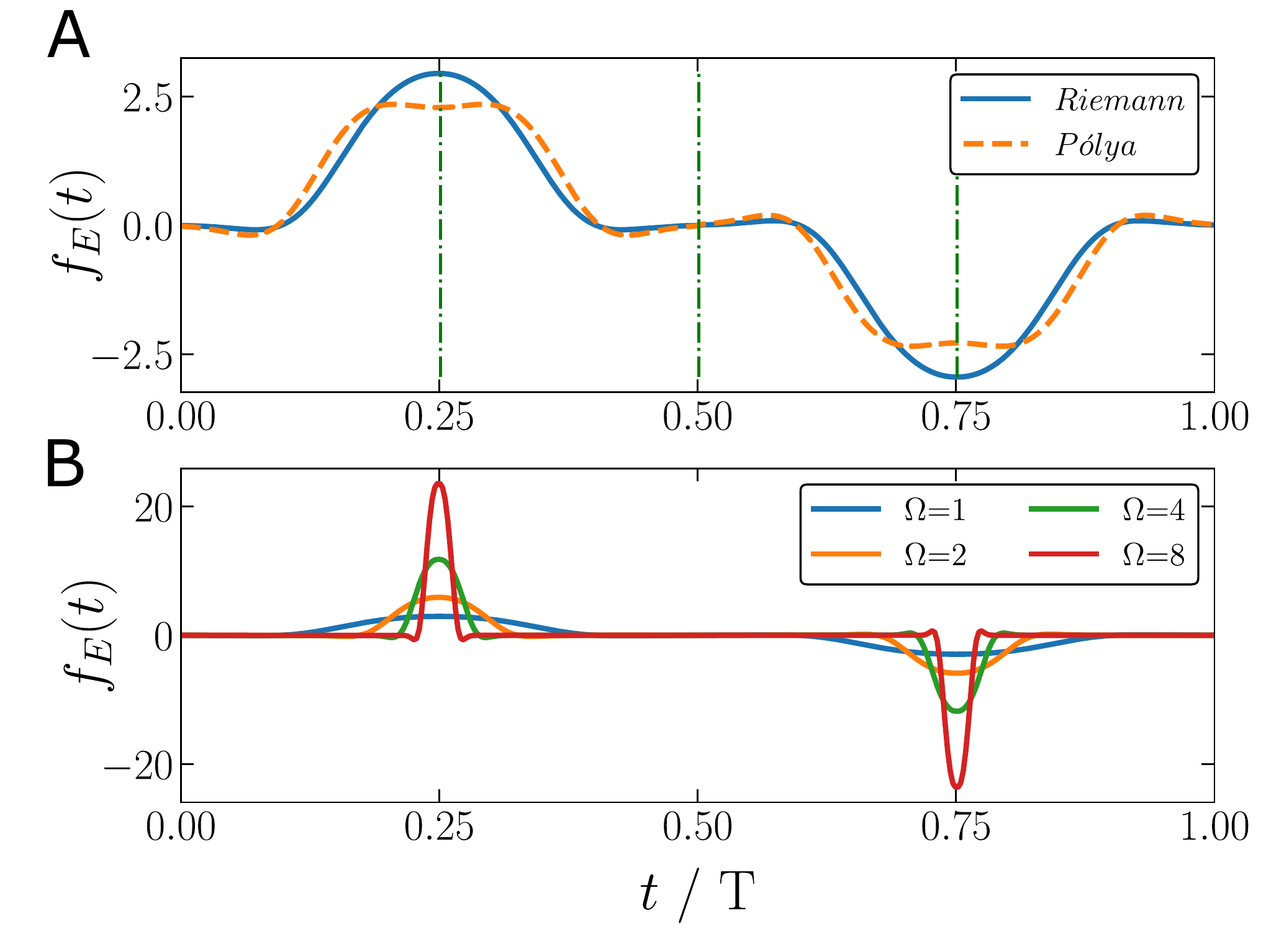}\caption{\label{fig:The-driving-potential} \textbf{Driving functions.} \textbf{(A)} The driving functions $f_{E}(t)$
for the Riemann $\varXi$ function (solid) and P\'olya's function (dashed)
for $E=4$. Four copies (separated by green-dashed lines) of the truncated $R_E(t)$ (Eq.\thinspace\ref{eq:Riemann's driving function-1}) and $P_E(t)$ functions (Eq.\thinspace\ref{eq:polya's driving function-1}) in the interval
$0\protect\leq t< \pi/2$ are joined together to obtain the continuous
periodic-driving function $f_{E}(t)$ with a vanishing time-average.
The driving period is $T$. \textbf{(B)} Pulse shape with different
$\Omega$. The period $T$ is fixed at a constant value ($2\pi$). The driving pulse
becomes narrower and higher as $\Omega$ increases. Note that a large $\Omega$
would ensure that the system is well within the high-frequency regime.
In our experiment, $\Omega$ is set to $8$.}
\end{figure}

\bigskip
\noindent\textbf{Dynamics of a Floquet system}
 
Because of their rich dynamics and flexibility, periodically driven systems
have been extensively used to explore different phenomena such as topological
insulators \citep{lindner2011floquet,grushin2014floquet}, non-equilibrium
dynamics \citep{xu2018measuring}, and time crystals \citep{sacha2015modeling,zhang2017observation}.
Generally, when a system is driven by the periodic Hamiltonian
$H(t)=H(t+T)$ (where T is the period), its dynamics can be effectively described by the Floquet
formalism \citep{floquet1883equations,shirley1965solution,grifoni1998driven},
i.e., the Floquet equation
\begin{equation}
(H(t)-i\hbar\frac{\partial}{\partial t})\left|\psi_{j}(t)\right\rangle =\epsilon_{j}\left|\psi_{j}(t)\right\rangle ,\label{eq:floquet Eq.}
\end{equation}
where the state $\left|\psi_{j}(t)\right\rangle $ is a Floquet state
satisfying the periodic condition  $\left|\psi_{j}(t)\right\rangle =\left|\psi_{j}(t+T)\right\rangle $,
and $\epsilon_{j}$ is the quasi-energy corresponding to the Floquet
state $\left|\psi_{j}(t)\right\rangle $. The Floquet state and the
quasi-energy are analogs of the eigenstate and the eigenvalue in
the time-independent case. The state of the driven system is the superposition
of the Floquet states, i.e. 
\begin{equation}
\left|\Psi(t)\right\rangle =\sum a_{j}\exp[-i\epsilon_{j}t]\left|\psi_{j}(t)\right\rangle ,\label{eq:Floquet state}
\end{equation}
where $a_{j}$ is the time-independent coefficient determined
by the initial state $\Psi(0)$. The Floquet states reflect the short-time
response of the system to the driving field within one period, whereas
the quasi-energies determine the long-term dynamics \citep{eckardt2009exploring,grifoni1998driven}.

We considered the periodic-driving Hamiltonian of a qubit system
as
\begin{equation}
H_{E}(t)=\hbar(-J\sigma_{x}+Jf_{E}(t)/2\:\sigma_{z}),\label{eq:hamiltonian}
\end{equation}
where $f_{E}(t)$ is the periodic driving field (with period $T$), i.e., $f_{E}(t)=f_{E}(t+T)$,
$E$ is the driving parameter, and $J$ is the tunneling frequency.
Generally, it is difficult to exactly solve the Floquet Eq.\thinspace\ref{eq:floquet Eq.}; however, in the strong-driving limit, where the frequency $\omega=2\pi/T$ dominates $J$, e.g., $\omega\gg J$, the Eq.\thinspace\ref{eq:floquet Eq.} can be solved by treating the time-independent part of
the Hamiltonian ($-J\sigma_{x}$) as a perturbation \citep{johansson2012qutip, creffield2003location}.
Based on the first-order expansion of $J$, the quasi-energies of
the driving qubit are given by $\epsilon_{\pm}=\pm\hbar|J_{eff}|$,
where 
\begin{equation}
J_{eff}=\frac{J}{T}\int_{0}^{T}dt\thinspace e^{-iF_{E}(t)} \label{eq:Jeffective}
\end{equation}
is the effective tunneling between the two Floquet states and 
\begin{equation}
F_{E}(t)=\int_{0}^{t}dt_1\thinspace f_{E}(t_1). \label{eq:F_f_relation}
\end{equation}
Therefore, we engineered effective tunneling via periodic driving.

\begin{figure*}[hbt]
\includegraphics[width=1\columnwidth]{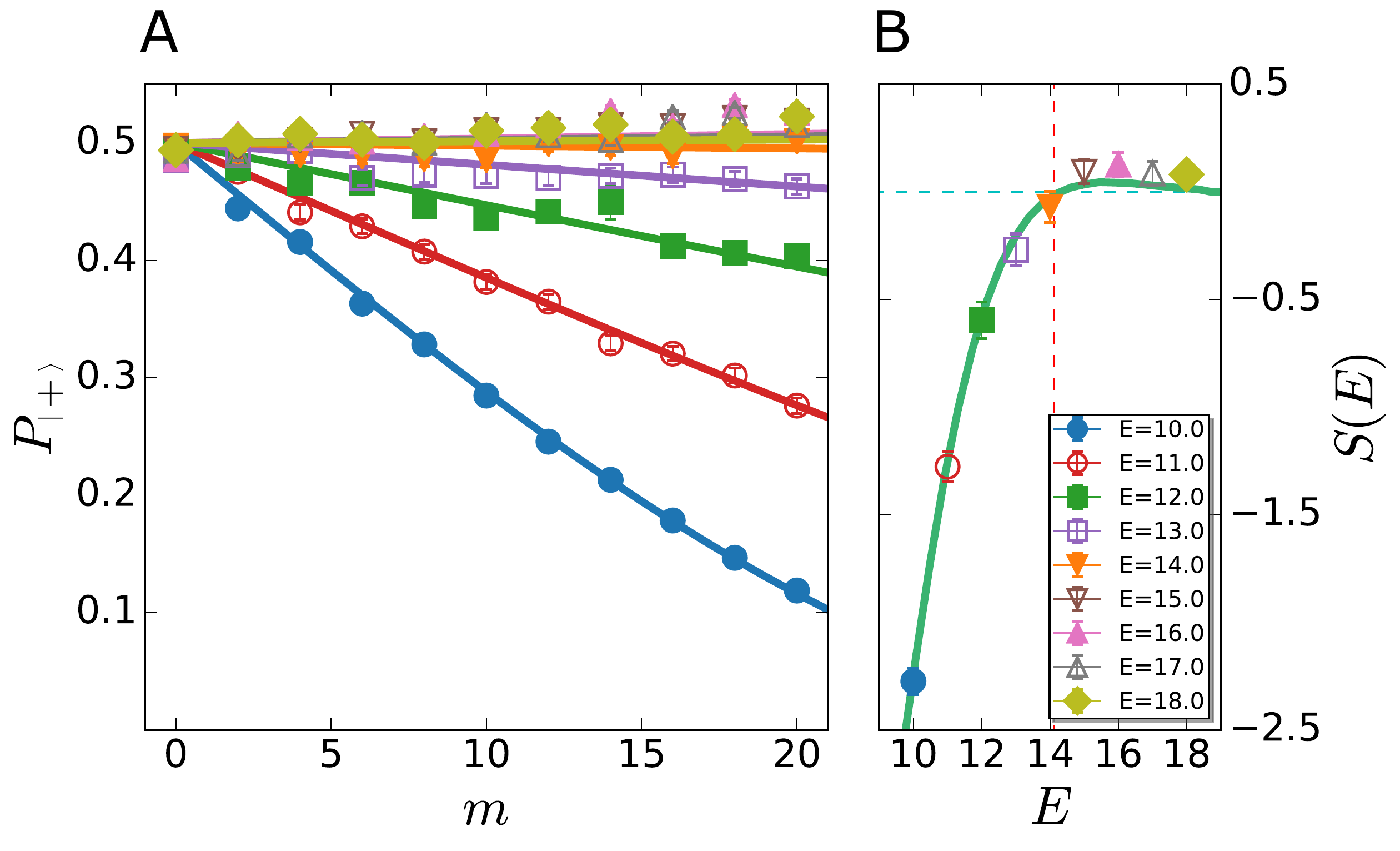}\includegraphics[width=1.05\columnwidth]{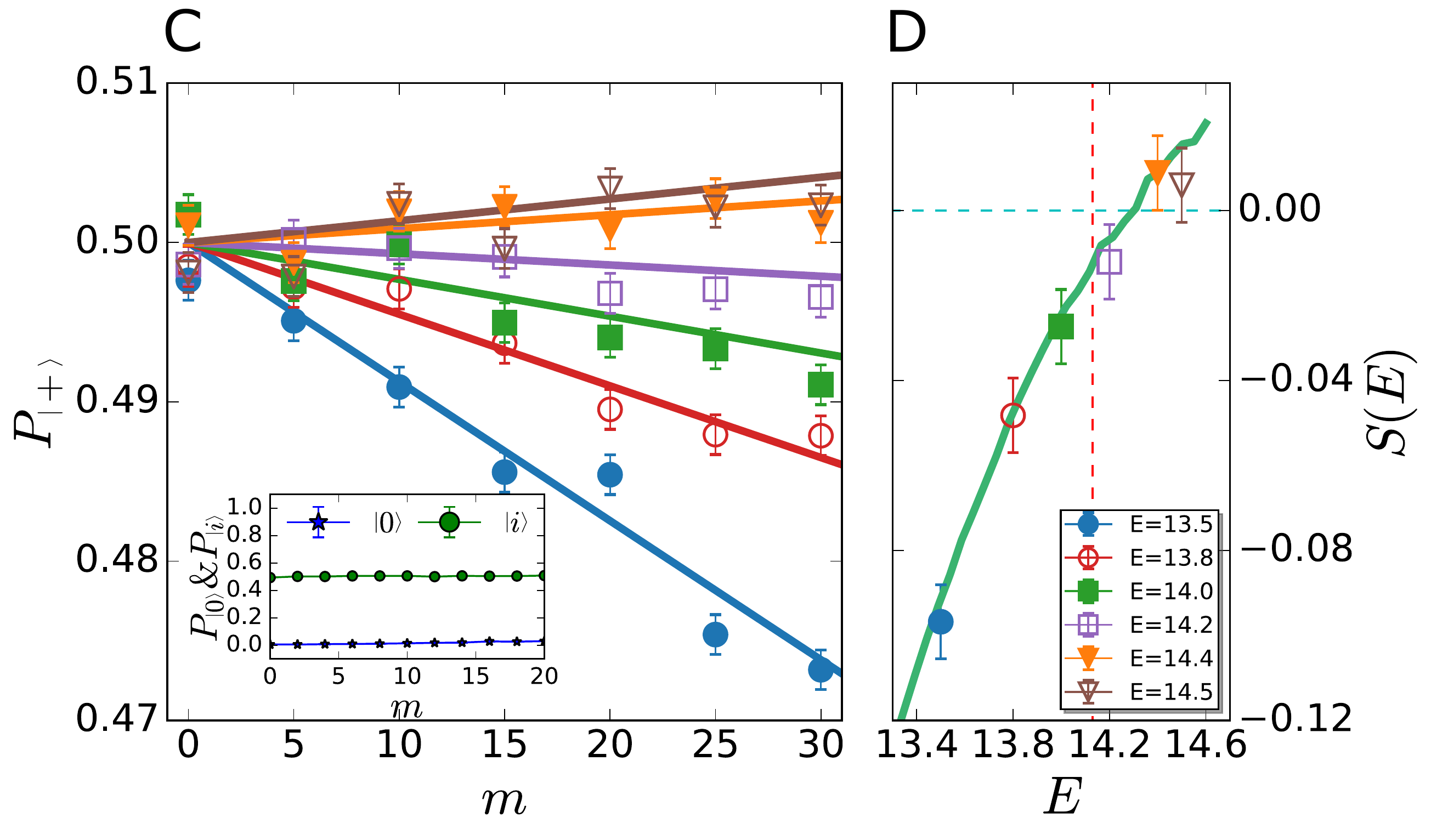}
\caption{\label{Fig:Riemann_first} \textbf{The dynamics of the probability $P_{\left|+\right\rangle}(E,t'=mT')$ with driving parameter $E$ near the first zero of the Riemann $\varXi(E)$ function.}
\textbf{(A)} In our experiment, the driving comprises up to 20 identical periods of $f_{E}(t)$. With different driving functions (characterized with different parameters $E$), the probability $P_{\left|+\right\rangle}$ at time $mT$ ($T=2\pi$ is the period) behaves differently. Under the tailored driving function $f_{E}(t)$, $\Re(J_{eff}(E))$ is proportional to $\varXi(E)$. Once the dynamics are frozen (i.e., CDT occurs), the associated $E$ corresponds to the zero of $\varXi(E)$. It is clear from (a) that the
evolution is nearly frozen (flat) when $E$ approaches 14 (CDT). \textbf{(B)}
The zero of the sum of the values of the residuals (SOR; the blue horizontal dashed line) lies between $E=14$ and $E=15$, indicating that
the first zero of the Riemann $\varXi$ function is between $E=14$
and $E=15$. The green solid line indicates the theoretical values of SOR. \textbf{(C)} For better precision, we obtain measurements from $E=13.5$
to $E=14.5$ with a step $\Delta E=0.1$ and drive the system up to $30$ periods. Each measurement is repeated 160,000 times. The inset in \textbf{(C)} depicts the measured probabilities $P_{\left|0\right\rangle}(E,t'=mT')$ and $P_{\left|i\right\rangle}(E,t'=mT')$ where $\left|i\right\rangle=(|0\rangle+i|1\rangle)$ with $E=14.0$. The SOR in \textbf{(D)} shows that the coherent destruction of tunneling (CDT) happens between $E=14.2$
and $E=14.4$, while the exact first zero of $\varXi(E)$
is $14.1347$ (the red vertical dashed line). }
\end{figure*}

\begin{figure*}[hbt]
\includegraphics[width=1\columnwidth]{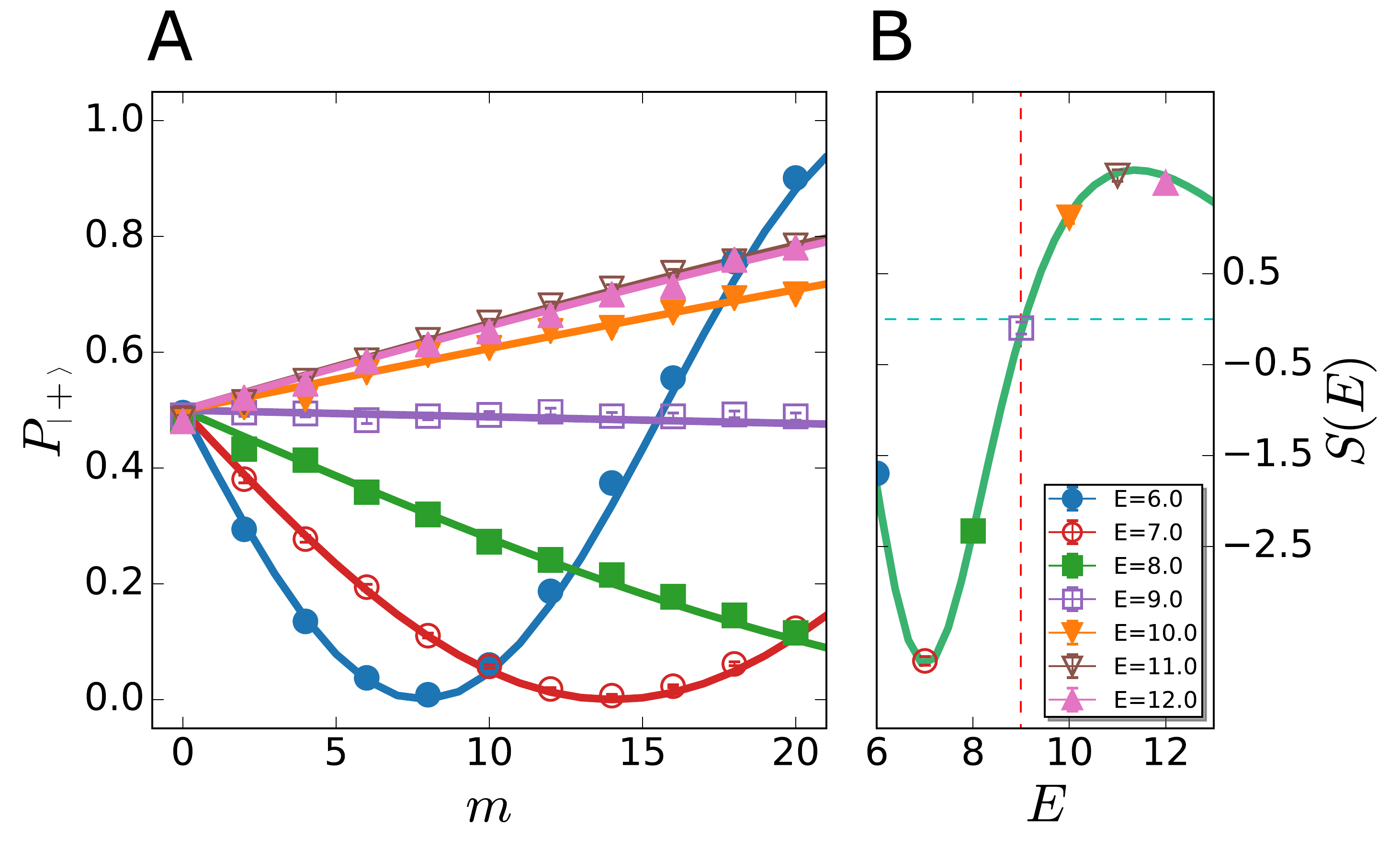}\includegraphics[width=1\columnwidth]{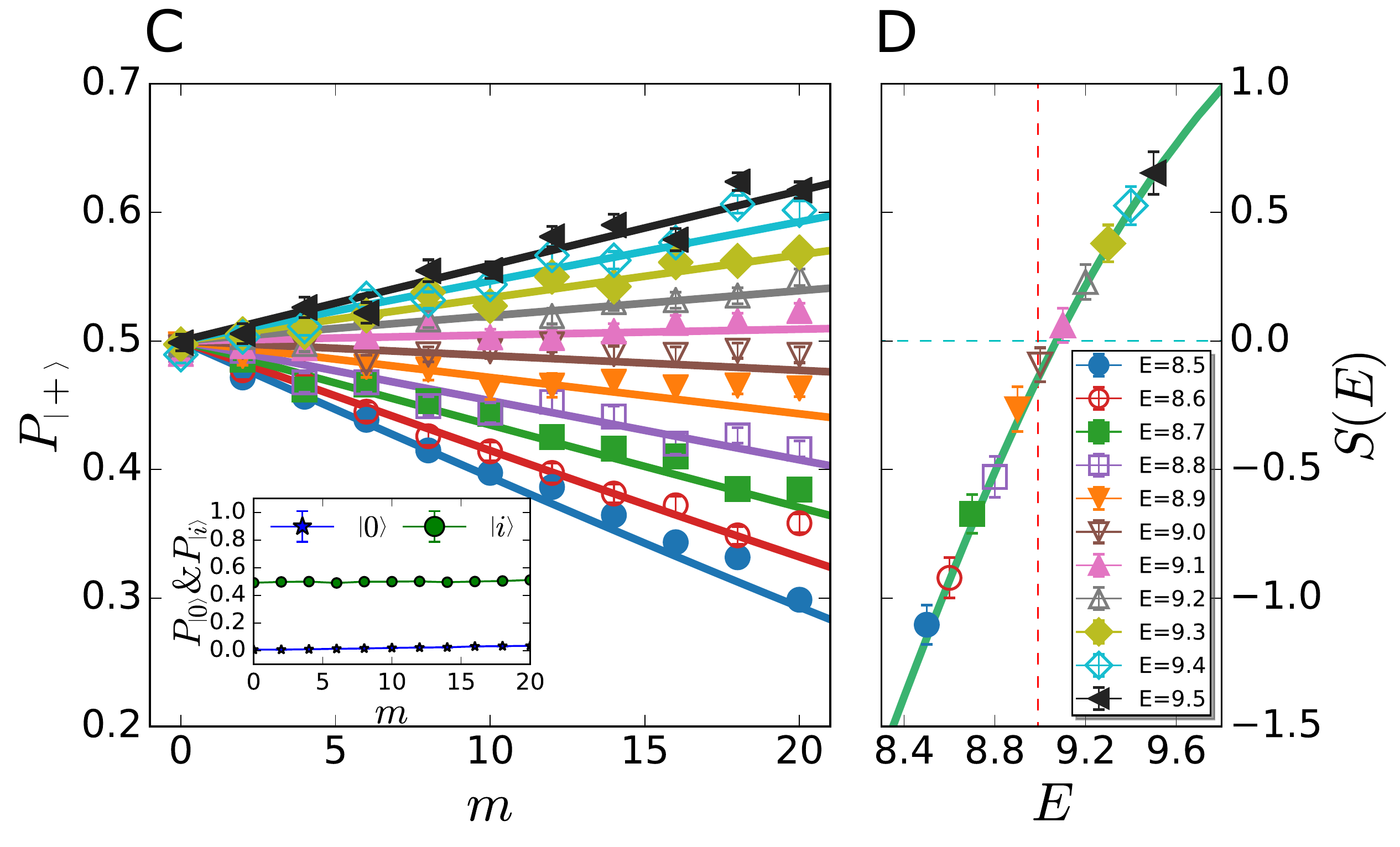}
\caption{\label{fig:Polya_first_rough} \textbf{The dynamics of the probability $P_{\left|+\right\rangle}(E,t'=mT')$ with driving parameter $E$ near the first zero
of P\'olya's function $\varXi^{*}(E)$.} \textbf{(A)} The coarse scan of $E$ from
6 to 12 explicitly indicates that the evolution is frozen around $E=9$.
\textbf{(B)} The measured (dots) and theoretical (green solid line) values of SOR. When $E=9$, SOR is nearly
zero (blue horizontal dashed line). A fine scan around $E=9.0$ with a step of 0.1 is depicted in
\textbf{(C)} and \textbf{(D)}. The inset in \textbf{(C)} depicts the measured probabilities $P_{\left|0\right\rangle}(E,t'=mT')$ and $P_{\left|i\right\rangle}(E,t'=mT')$ with $E=9.0$. It can be inferred that the zero of
the SOR as well as P\'olya's function $\varXi^{*}(E)$ lies between 9.0
and 9.1. The deviation from the exact zero of 8.993 (the red vertical dashed line) is because of the limited $\Omega$.}
\end{figure*}

\bigskip
\noindent\textbf{Periodic-driving function for the Riemann $\varXi$ function and P\'olya's function}

As shown in the methods section, if
the driving field $f_{E}(t)$ in the Eq.\thinspace\ref{eq:hamiltonian} is carefully designed, the effective tunneling $\Re(J_{eff}(E))$ can be proportional to the Riemann $\varXi$ function, i.e., $\Re(J_{eff}(E))\propto\varXi(E)$. The Riemann $\varXi$ function is $\varXi(E)=\frac{1}{2}s(s-1)\Gamma(\frac{s}{2})\pi^{-s/2}\zeta(s)$, whose zeros coincide with the non-trivial zeros of the Riemann $\zeta(s)$ function. Consequently, when $E$ is the zero of $\varXi(E)$, $J_{eff}(E)$ will vanish, indicating that the dynamics of the system are frozen. This phenomenon is the so-called CDT \citep{grossmann1991coherent,grifoni1998driven}. Therefore, the zeros of the Riemann $\varXi$ function which coincide with the non trivial zeros of $\zeta(s)$ can be obtained by detecting the degeneracy of the quasi-energies as the parameter $E$ varies. 

To achieve the desired effective tunneling, the driving function in one period is formed of four segments: the first segment ($0\leq t\leq \pi/2$) can be described as $f_E(t)=R_{E}(\pi/2-t)$ (Eq.\thinspace \ref{eq:Riemann's driving function-1} in the Methods Section); the second segment ($\pi/2 \leq t\leq \pi$) is obtained by reflection transformation of the first segment along $t=\pi/2$, i.e., $f_E(t)=f_E(\pi-t)$; and the third and fourth segments ($\pi \leq t\leq 2\pi$) can be constructed as $f_E(t)=-f_E(2\pi-t)$. The driving function in the whole period is shown in Fig.\ \ref{fig:The-driving-potential}A. Note that this function is continuous, has a time-average of zero, and satisfies the required parity condition (details in the Methods section). 

\begin{figure}[hbt]
\includegraphics[width=1\columnwidth]{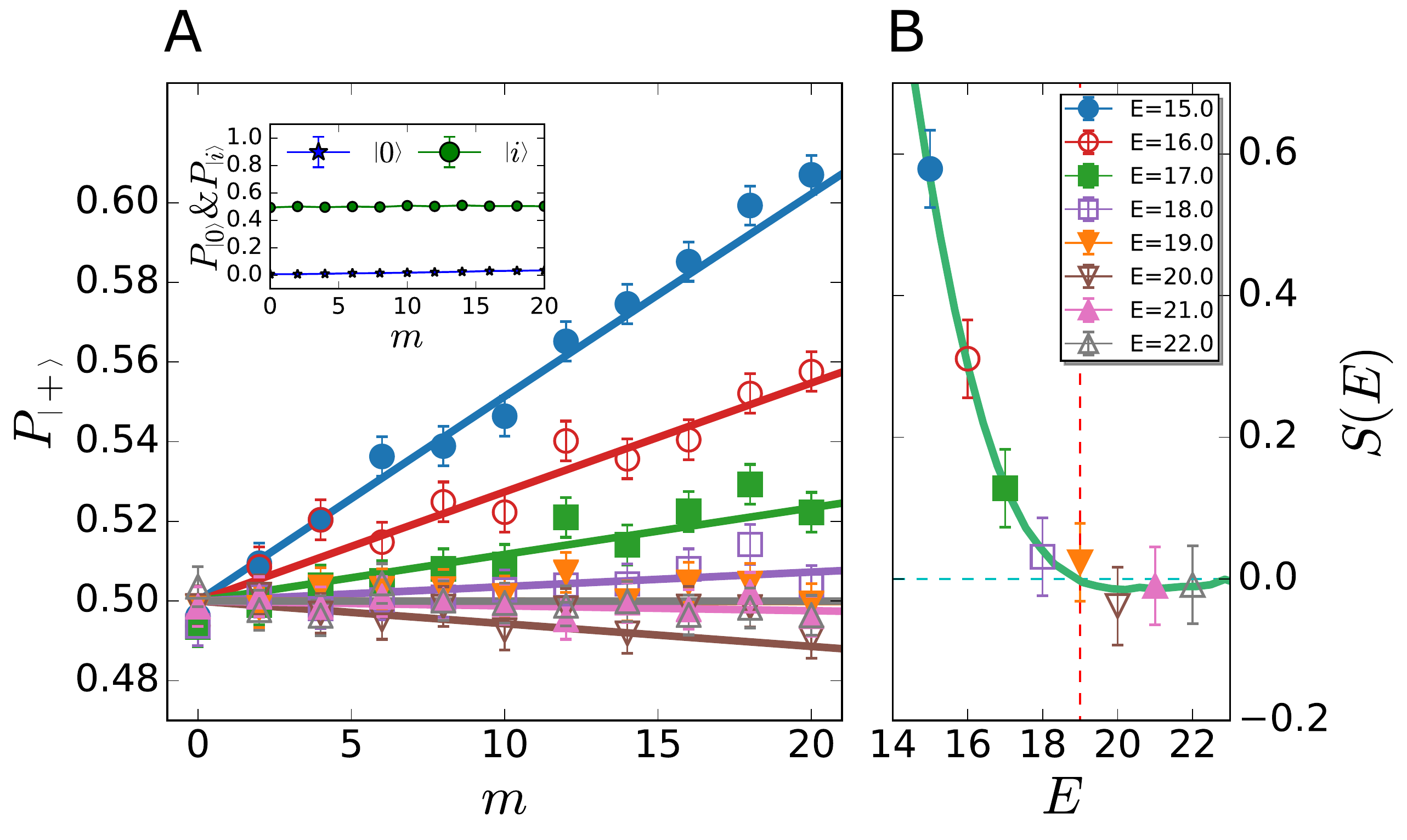}
\caption{\label{fig:Polya_second} \textbf{The dynamics of the probability $P_{\left|+\right\rangle}(E,t'=mT')$ with a driving function near the second zero
of P\'olya's function $\varXi^{*}(E)$.} \textbf{(A)} When $E$ increases from $15$
to $22$, the second CDT and therefore the second
zero point of $\varXi^{*}(E)$ appears between $E=18$ and $E=19$. The inset in (a) depicts the measured probabilities $P_{\left|0\right\rangle}(E,t'=mT')$ and $P_{\left|i\right\rangle}(E,t'=mT')$ where $\left|i\right\rangle=(|0\rangle+i|1\rangle)$ when $E=19.0$. \textbf{(B)} The SOR values for each $E$ vanish between $E=18$ and $E=19$, which agrees well with the exact zero, i.e., $18.996$. }
\end{figure}

Similarly, P\'olya's function 
$\Xi^{*}(E)$, which can be viewed as a smoothed version of the Riemann
 $\varXi(E)$ function, can be proportional to the effective tunneling of a certain engineered driving function $P_E(t)$ (the form of the function is shown in Fig.\ \ref{fig:The-driving-potential}A and Eq.\thinspace \ref{eq:polya's driving function-1} in the Methods section). Its zeros can be obtained by detecting the degeneracy of quasi-energies by scanning the parameter $E$ in the driving field.

To ensure the validity of the previous statement, two contradictory conditions need to be guaranteed: a) the high-frequency condition in Eq.\thinspace\ref{eq:hamiltonian}, ($\omega=2\pi/T \gg J$), which validates the perturbation expansion in the Floquet formalism, and b) the large-$T$ condition, which guarantees that the effective tunneling converges well to the desired function. Actually, the driving functions $R_{E}(t)$ and $P_E(t)$ decay rapidly and the function cuts off at $T_0=\pi/2$ is enough to ensure convergence. To further satisfy the high-frequency condition, we fixed the period of the driving function to $2\pi$ and re-scaled the driving field $f_E(t)$ as $\Omega f(\Omega t)$, where $\Omega>1$. The re-scaled driving field is shown in Fig.\ \ref{fig:The-driving-potential}B. For a larger $\Omega$, the driving pulses get narrower and higher and the quasi-energy spectrum of the system will approach the high-frequency limit. Thus, with a larger $\Omega$ value, the results are more precise. Note that, for our experiment, $\Omega=8$.

\bigskip
\noindent\textbf{Experimental setup}

To demonstrate the previous theories, the ion-trap simulator is the ideal setup. In the experiment, the two-level system is encoded in a trapped ion. The qubit to be driven is encoded in the 12.6-GHz hyper-fine clock transition $\left|0\right\rangle \equiv{}^{2}S_{1/2}\left|F=0,\thinspace m_{F}=0\right\rangle $
and $\left|1\right\rangle \equiv{}^{2}S_{1/2}\left|F=1,\thinspace m_{F}=0\right\rangle $
in a single $^{171}\mathrm{Yb}^{+}$ ion confined in an RF trap \citep{cui2016experimental}.
This transition is first-order magnetic-field insensitive and a 
long coherence time of 10 min has been observed using sympathetic
cooling and dynamical decoupling \citep{wang2017single}. The long coherence of this qubit is critical to observe the CDT (corresponding to the zeros of the corresponding functions), for which many driving cycles are necessary.

After 1 ms of Doppler cooling, the qubit is initialized
to the ground state $\left|0\right\rangle $ with a probability of
99.9\%. Then, a modulated microwave-driving pulse $B(t')=B_{0}\cos(\omega_{01}t'+J\phi(t=Jt'))$ is generated using a programmable
arbitrary waveform generator, where $\omega_{01}$ is the qubit transition frequency and $t'=t/J$ is the real time in units of $s$ and $\phi(t)=F(t)/2$. Moving to the rotating frame of the
microwave using the unitary transformation $e^{-i\frac{\omega_{01}t'+J\phi(Jt')}{2}\sigma_{z}}$
and applying the rotating-wave approximation, we obtain the interaction
Hamiltonian Eq.\thinspace\ref{eq:hamiltonian} between the microwave and
the atom \citep{de2016estimation}. Note that the tunneling frequency
$J$ is set to $(2\pi\text{)}4$ kHz, which corresponds to a Rabi
time of $250$ $\mu$s. Moreover, the re-scaled parameter $\Omega$ is set to $8$, and the qubit is then driven for time $t'=mT'$, where $m$ is an integer
and $T$ is the fundamental driving period $T'=2\pi/J$. 
Subsequently, a resonant $\frac{\pi}{2}$ microwave pulse is applied
to transfer the measurement basis to $\left|+\right\rangle=\frac{1}{\sqrt{2}}(\left|0\right\rangle+\left|1\right\rangle)$.
Finally, a $400$-$\mu$s pulse of a $369.5$-nm laser is immediately
used for fluorescence detection. When
more than one photon is detected, the measurement result is noted as 1; otherwise, it is noted as 0. Many repetitions are performed to obtain the $P_{\left|+\right\rangle}$. Once we determine that $P_{\left|+\right\rangle}$ does not change with m at some $E$, $P_{\left|0\right\rangle}$ and $P_{\left|i\right\rangle}$ are measured to confirm the occurrence of CDT, where $\left|i\right\rangle=\frac{1}{\sqrt{2}}(\left|0\right\rangle+i\left|1\right\rangle)$.

\bigskip
\noindent\textbf{Experimental results of the Riemann $\varXi(E)$ function}

We first detect the zeros of the Riemann $\varXi(E)$ function. One period of the fundamental driving function $f_{E}(t)$ is depicted in Fig.\ \ref{fig:The-driving-potential}.
To detect the CDT, we scan the driving parameter $E$ and can measure the probabilities of the state on different bases at $t'=mT'$ (where $m$ is an integer). The CDT occurs at a certain $E$ where the measured probabilities are constant for all $m$. In the experiment, for better precision, we use the probability $P_{|+\rangle}$ projected onto the basis $|+\rangle$ as the primary indicator of CDT (the details of the selection of the projection basis can be found in the Methods section). 

To identify the $E$ corresponding to the CDT, $E$ is first scanned between 10 and 18 with a step size $\Delta E=1$. In order to quantitatively measure the deviation between the measured evolution and the CDT, we define the SOR as $S(T'_{m},E)=\sum_{m}P_{|+\rangle}(T'_{m},E)-P_{CDT}(T'_{m})$, where $P_{|+\rangle}(T'_{m},E)$ denotes the probability that the state is measured to be in the state $|+\rangle$ at time $T'_{m}$ and $P_{CDT}(T'_{m})$ is the probability when CDT occurs. $P_{CDT}(T'_{m})$ is determined by the prepared state and the measurement basis, e.g., in our current experiment, $P_{CDT}(T'_{m})=0.5$. The SOR, noted as $S(E)$, depends upon $E$ and can be positive or negative. In particular, when CDT occurs, $S(E)$ will be zero. Actually, the best quantity to characterize the CDT is the quasi-energy. In the Method section, we give the method to extract quasi-energy $\epsilon(E)$ from the evolution of $P_{|+\rangle}(T'_{m},E)$, i.e., $P_{\left|+\right\rangle}=1/2-\sin(2\epsilon(E) t')/2$. However, as $E$ is approaching the zero point, the evolution becomes very slow and it becomes difficult to extract $\epsilon(E)$ with only 20 periods of driving.

As can be seen in Fig.\ \ref{Fig:Riemann_first}A, when $E$ approaches $14$, the probability $P_{|+\rangle}$ on the basis $\left|+\right\rangle $ is stable and barely changes
from 0.5, indicating that the CDT happens near $E=14$. Note that $S(E=14)<0$ and $S(E=15)>0$, which suggests that the CDT will appear between $E=14$ and $15$.

To determine the best value of $E$ at which the CDT appears, we utilized a smaller $\Delta E$ and scanned $E$ from $13.5$ to $14.5$. The experimental results are depicted in Fig.\ \ref{Fig:Riemann_first}C and Fig.\ \ref{Fig:Riemann_first}D.
Because $S(E=14.2)<0$ and $S(E=14.4)>0$, we inferred that the zero of $S(E)$ turns out to be between $14.2$ and $14.4$. Based on this result, it can inferred that the zero of $J_{eff}(E)$ and $\varXi(E)$ lies between $14.2$ and $14.4$, which is very close to the exact first zero of Riemann zeta function $\zeta(1/2+iE)$, i.e., $E=14.1347$. As shown below, when $\Omega$ is higher, the result will be more precise.

The experiment is repeated $40,000$ times in Fig.\ \ref{Fig:Riemann_first}A and $160,000$ times in Fig.\ \ref{Fig:Riemann_first}C. The error bars in Fig.\ \ref{Fig:Riemann_first}A and Fig.\ \ref{Fig:Riemann_first}C
indicate the statistical error within one standard deviation. The error bars of SOR in Fig.\ \ref{Fig:Riemann_first}B and Fig.\ \ref{Fig:Riemann_first}D
are the sums of the statistical errors of the corresponding $m$
data points $p_{0}(m,E)$ in Fig.\ \ref{Fig:Riemann_first}A and
Fig.\ \ref{Fig:Riemann_first}C, respectively. 

\bigskip
\noindent\textbf{Experimental results of P\'olya's function $\varXi^{*}(E)$}

We will now focus on identifying zeros of P\'olya's function. Similar to the Riemann $\varXi$ function situation, we first scan the parameter $E$ with a step size $\Delta E=1$. The experimental results are shown in Fig.\ \ref{fig:Polya_first_rough}A
and Fig.\ \ref{fig:Polya_first_rough}B. The first zero of $S(E)$ falls around $E=9$. To obtain a more precise value of $E$, we scanned $E$ from $8.5$ to $9.5$ with a step size of $\Delta E=0.1$ and the results are shown in
Fig.\ \ref{fig:Polya_first_rough}C and Fig.\ \ref{fig:Polya_first_rough}D.
Note that $P_{|+\rangle}$ remains unchanged when $E$ is between $9.0$ and
$9.1$. Therefore, the zeros of $J_{eff}(E)$
and $\varXi^{*}(E)$ are measured to be between $9.0$ and $9.1$,
which is close to the exact first zero of P\'olya's function
$\varXi^{*}(E)$, i.e., $E=8.993$. Furthermore, we measured the second
zero of P\'olya's function, which is near $19$, as shown
in Fig.\ \ref{fig:Polya_second}. For each point, the experiment was repeated $40,000$ times. 

\begin{figure}[hbt]
\includegraphics[width=1\linewidth]{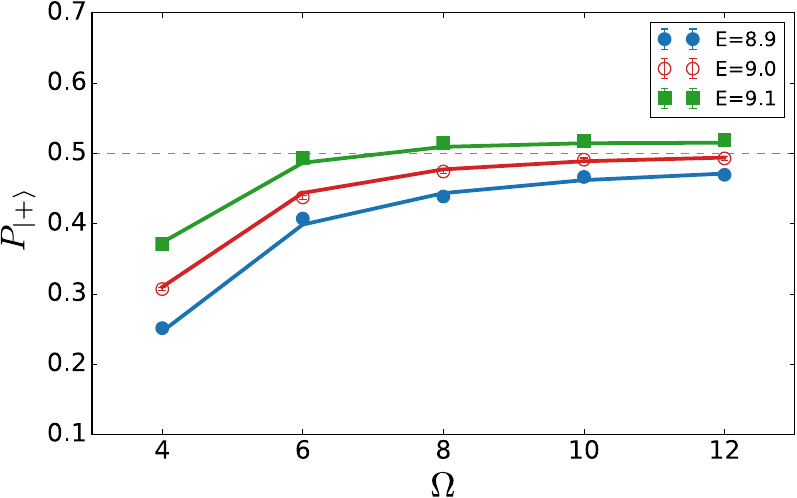}

\caption{\label{fig:3} \textbf{The dynamics of the probability $P_{\left|+\right\rangle}$ versus the scale factor $\Omega$ near P\'olya's function's first zero.} The larger value of $\Omega$ indicates a better high-frequency approximation and a higher precision of the zeros. The system undergoes CDT with the driving parameter $E$ above $9.1$ when $\Omega=6$; however, when the parameter $\Omega$ increases to $8$, CDT occurs with the parameter $E$ between $9.1$ and $9.0$. If $\Omega$ further increases
to $12$, CDT appears with the parameter $E$ remaining very close to $9.0$, which is closer to the exact result of $E=8.993$. Thus, if $\Omega$ keeps increasing, the CDT will theoretically approach the exact zero of the $\varXi^{*}(E)$ function. Moreover, the results show that the limited $\Omega$ is the dominant reason for the deviation in zeros.} 
\end{figure}

The experimental result and the theoretical prediction in Fig.\ \ref{fig:Polya_first_rough}D
are slightly larger than the exact zero $8.993$ 
because the scale factor $\Omega$ is not large enough to satisfy the
high-frequency limit. To demonstrate that the zeros can be better approached with a higher value of $\Omega$, the behaviors of $P_{0}(t'=20T',E)$ at $E=8.9$, $9.0$
and $9.1$ with varying $\Omega$ are shown in Fig.\ \ref{fig:3}.
The smaller $\Omega$ ($\Omega < 6$ ) introduces a relatively
larger deviation from the actual zero of P\'olya's function $\varXi^{*}(E)$, e.g., when $\Omega\leq6$, CDT occurs with the driving parameter $E > 9.1$. Moreover, CDT appears with the parameter $E<9.1$ when $\Omega\geq8 $ and approaches the exact value as $\Omega$ increases. However, CDT is much closer to $E=9.0$ when $\Omega=12$; therefore, larger values of $\Omega$ indicate that the zeros have better precision. However, from an experimental point of view, a larger $\Omega$ indicates more driving periods to distinguish the evolutions of the state with different $E$. This will be limited by the coherence time and the stability of the system. Therefore, under the trade off, we selected $\Omega=8$ in our current experiment. Actually, for the Riemann $\varXi(E)$ function, it is more challenging to achieve larger $\Omega$ than it is with P\'olya's function because of its smaller quasi-energies. 

\bigskip

In conclusion, we have periodically driven a two-level quantum system and identified that the special values of a control parameter where the CDT phenomena happens coincidents with the first non-trivial zero of the Riemann zeta function and the first two zaros of P\'olya's fake zeta function. Although this is not a direct realization of  the original P\'olya--Hilbert conjecture, it supplies an indirect observation of the Riemann zeros in a quantum system for the first time. By observation of the vanishing of the quasi-energies of a periodically driven system, the Riemann zeros can be indentified presicely. This work opens a new method to pursue the secret of
the PHC in a real physical system, as well as the famous Riemann hypothesis. In the future, this method can be extended to indentify more zeros
by improving the robustness and detection efficiency of the platform or by using a more efficient driving function for which the quasi-energies decay more
slowly \citep{Berry2012Riemann}.

\bigskip
\noindent\textbf{METHODS}

\noindent\textbf{Driving function}
 
We briefly introduce how driving functions can be obtained \citep{creffield2015finding}. On one hand, the effective Hamiltonian of the driving system with the Hamiltonian (\ref{eq:hamiltonian}) under the first-order perturbation is
\begin{myequation}
H_{eff}=J_{eff}\sigma_{x}.  \label{eq:H_effective}
\end{myequation}
The quasi-energies of the driving qubit are given by $\epsilon_{\pm}=\pm|J_{eff}|$, where $J_{eff}$ is determined by Eq.\thinspace\ref{eq:Jeffective}. The real part of  $J_{eff}(E)$ is 
\begin{myequation}
\Re(J_{eff}(E))=\int_{0}^{T}dt\cos F_{E}(t).  \label{eq:real_of Jeff}
\end{myequation}

On the other hand, the Riemann $\varXi(E)$ function can be written as \citep{Polya1926Bemerkung} 
\begin{myequation}
\varXi(E)=\int_{0}^{\infty}dt\Phi(t)\cos(Et/2), \label{eq:Riemann_Xi_function_integration}
\end{myequation}
where $\Phi(t)=2\pi e^{5t/4}\sum_{n=1}^{\infty}(2\pi e^{t}n^{2}-3)n^{2}e^{-\pi n^{2}e^{t}}$ (in this work, we used the first 100 terms).  The aim is to find a $F_{E}(t)$ such that the real part of $J_{eff}(E)$ is propotional to the Riemann $\varXi(E)$ function, i.e.,  
\begin{myequation}
\alpha\int_{0}^{\infty}dt\Phi(t)\cos(Et/2)=\int_{0}^{T}dt\cos F_{E}(t), \label{eq:propotional_Eq.}
\end{myequation}
where $\alpha$ is the constant of proportionality between  $J_{eff}(E)$ and the zeta function. For T sufficiently large, we can replace the upper limit of integration on the left  to obtain
\begin{myequation}
\alpha\int_{0}^{T}dt\Phi(t)\cos(Et/2)=\int_{0}^{T}dt\cos F_{E}(t). \label{eq:propotional_T_Eq.}
\end{myequation}
Note that this relation must hold for all sufficiently large values of T (not just one particular value of T). As a consequence we can deduce that
\begin{myequation}
F_{E}(t)=\cos^{-1}[\alpha\Phi(t)\cos(Et/2)], \label{eq:Big_F}
\end{myequation}
Boundary conditions require $F(t=0)=0$, and thus $\alpha \Phi(0)=1$ and $\alpha=1/\Phi(0)$. This allows us to write $F(t)$ as
\begin{myequation}
F_{E}(t) = \cos^{-1}[\Phi(t) / \Phi(0)  \cos(E t/2) ], \label{eq:Big_F2}
\end{myequation}
Having obtained this result, we can now choose a specific value for T. Because $\Phi(t)$ decays rapidly, the integration on the right hand side can be cut at $T=\pi/2$. 
Thus we choose $T=\pi/2$. By this choice of $F_{E}(t)$ the real part of $J_{eff}(E)$ becomes proportional to $\varXi(E)$. If $E$ is the zero of Riemann zeta function,  it will be also the zero of Rieman $\varXi(E)$ function and the  $J_{eff}(E)$ will vanish. 

From Eq.\thinspace\ref{eq:F_f_relation} and Eq.\thinspace\ref{eq:Big_F2}, the driving function for the Riemann $\varXi(E)$ function can be directly obtained as follows:
\begin{myequation}
R_{E}(t)=-\frac{[\Phi'(t)\thinspace\cos(Et/2)-(E/2)\thinspace\Phi(t)\thinspace\sin(Et/2)]}{\sqrt{\Phi^{2}(0)-[\Phi(t)\cos(Et/2)]^{2}}}, \label{eq:Riemann's driving function-1}
\end{myequation}
where $t=Jt'\in[0,\pi/2]$ represents a dimensionless parameter \citep{creffield2015finding}. 

The same method can be used on P\'olya's function, 
\begin{myequation}
\Xi^{*}(E)=4\pi^{2}[K_{a+iE/2}(x)+K_{a-iE/2}(x)], \label{eq:polya's function}
\end{myequation}
where $x=2\pi$, $a=9/4$ and $K_{\beta}(t)$ is the modified K- Bessel function, $K_{\beta}(x)=\int_{0}^{\infty}dt\,\cosh(\beta t)e^{-x\cosh\,t}$. Thus 
\begin{myequation}
\Xi^{*}(E) = \int_{0}^{\infty}dt\,\alpha\cosh(at)e^{-2\pi\cosh\,t}\cos(Et/2). \label{eq:polya's  function-2}
\end{myequation} Replacing the upper limit with $\pi/2$ will only lead to an negligible relative error on the order of $10^{-7}$.
From $\Re(J_{eff}(E))\propto\Xi^{*}(E)$, the driving function can be obtained as follows:
\begin{myequation}
P_E(t)=-\frac{\phi(t)(a\thinspace\tanh\thinspace at-2\pi\thinspace\sinh\thinspace t-E/2\thinspace\tan(Et/2))}{\sqrt{1-\phi(t)^{2}}}, \label{eq:polya's driving function-1}
\end{myequation}
where $a=9/4$ and $t=Jt'\in[0,\pi/2]$.

Both $R_{E}(t)$ and $P_{E}(t)$ are truncated at $\pi/2$. The driving function $f_{E}(t)$ constructed by directly repeating $R_{E}(t)$ or $P_{E}(t)$ $m$ times has the following limitations: i) it is discontinuous;
ii) its average over one period does not vanish, which may heat the cold atom; iii) the periodic driving
field does not have odd parity, which indicates that the quasi-energies may
form a broad avoided crossing according to the von Neumann--Wigner theorem \citep{von1993verhalten}. To ensure the driving field $f_{E}(t)$ satisfies the parity requirement $\mathcal{P}:\ x\rightarrow-x,\ t\rightarrow t+T/2$, we
joined together four copies of $R_E(t)$ or $P_E(t)$ as shown in Fig.\ \ref{fig:The-driving-potential}A with the total period $T=2\pi$ (in the experimental setting, the period $T'$ is $2\pi/J$). Note that these modified driving functions, which are introduced in the main text, can overcome all of the above problems and remain experimentally achievable. 

\bigskip
\noindent\textbf{Measurement-bases selection}
 
Generally, CDT can be detected by projecting the states at time $mT$ onto any basis; however, the sensitivity of the probability against parameter $E$ is strongly dependent upon the measurement basis. Here, we show how to select the measurement basis such that the CDT can be observed in a reasonable driving time and at a higher precision. 

In our experiment, the initial state in Eq.\thinspace\ref{eq:Floquet state} is uniformly prepared in $\left|0\right\rangle$. We will now compare three often-used measurement bases: $\left|0\right\rangle$ (basis for $\sigma_z$), $\left|+\right\rangle$ (basis for $\sigma_x$), and $\left|i\right\rangle=\frac{1}{\sqrt{2}}(\left|0\right\rangle\ + i\left|1\right\rangle)$ (basis for $\sigma_y$).

1) If we project the system on the basis $\left|0\right\rangle$, the probability at time $t'=mT'$ with parameter $E$ will be $P_{\left|0\right\rangle}(E,t')=1/2+ \cos(2\epsilon(E)t')/2$, where $\epsilon(E)$ is the quasi-energy defined in Eq.\thinspace\ref{eq:floquet Eq.}. For two close parameters $E_{1}$ and
$E_{2}$, the difference of their probability is $\triangle P_{\left|0\right\rangle}=P_{\left|0\right\rangle}(E_{1},t')-P_{\left|0\right\rangle}(E_{2},t')=t'^{2}(\epsilon^{2}(E_{1})-\epsilon^{2}(E_{2}))+\mathcal{O}((t'\epsilon)^{2k})$,
where $\mathcal{O}((t'\epsilon)^{2k})$ is the higher even-order terms.

2) If we project the state on the bases $\left|+\right\rangle $,
the probability is $P_{\left|+\right\rangle}=1/2-\sin(2\epsilon(E) t')/2$. Similarly, the difference of the probability with two close parameters $E_{1}$ and $E_{2}$ is 
\begin{myequation}
\triangle P_{\left|+\right\rangle}=t'(\epsilon(E_{1})-\epsilon(E_{2}))+\mathcal{O}((t'\epsilon)^{2k+1}),\label{eq:deltaP}
\end{myequation}
where $\mathcal{O}((t'\epsilon)^{2k+1})$ gives the higher odd-order terms.

3) If we project the system on the basis $\left|i\right\rangle$, the probability $ P_{\left|i\right\rangle}$ will sinusoidally oscillate near 0.5, with the amplitude $A$ being much smaller than $1$ ($A\simeq0.005$ for Riemann function ). The probability difference will be considerably small to be measured experimentally.

The quantity $t'\epsilon$ is much smaller than $1$ when $E$ approaches CDT, which indicates $\triangle P_{\left|0\right\rangle}, \triangle P_{\left|i\right\rangle}\ll\triangle P_{\left|+\right\rangle}$. 
To obtain precise values of zeros of the Riemann $\varXi(E)$ and P\'olya's functions, we need to separate $P_{\left|+\right\rangle}(E_{1},t')$ from $P_{\left|+\right\rangle}(E_{2},t')$ as far as possible, where $E_1$ and $E_2$ are close to each other. The previous analysis shows that $P_{|+\rangle}$ is much more sensitive against the parameter $E$ then $P_{|0\rangle}$ and $P_{|i\rangle}$. Therefore, in our experiment, we selected $P_{|+\rangle}$ as our primary CDT indicator. 
Furthermore, Eq.\thinspace\ref{eq:deltaP} indicates that if we want to have a larger $\triangle P_{\left|+\right\rangle}$, a longer driving time $t'$ will be expected. In the experiment, because of the SPAM error (99.5\% in our system), the best distinguishable $\triangle P_{\left|+\right\rangle}$ is around $0.005$. If CDT occurs at $E_{2}$ and $\Delta E=0.1$, $\epsilon(E_{1})-\epsilon(E_{2})$
will be of the order of $(5\times10^{-4})/T'$, which indicates that the driving time
$t$ should be at least $10T'$ according to Eq.\thinspace\ref{eq:deltaP}. Moreover, the coherence time should be longer than $2,000$ periods (\textasciitilde $500$ ms in our case). However, to identify the second zero of Riemann function, $t'$ should be longer than $1,000T'$ and the coherence time needs to be over $10^{5}T'$ ($\sim30$ seconds in our case). 

\bigskip
\noindent\textbf{Observation of CDT in a trapped-ion qubit} 

Unlike the original proposal that $J_{eff}$ can be directly measured by
the free-expansion rate of a Gaussian wave packet of cold atoms \citep{della2007visualization,lignier2007dynamical,creffield2015finding},
we detect the vanishing of the effective tunneling $J_{eff}$ using the frozen dynamics in our trapped ion. Generally, the system is in the state described by Eq.\thinspace\ref{eq:Floquet state} whose evolution depends only upon the quasi-energies.
When CDT occurs, the vanishing quasi-energies in Eq.\thinspace\ref{eq:Floquet state}
merely contribute a global phase to the after-integer periods, i.e., $\left|\Psi(t+kT)\right\rangle =e^{i\varphi}\left|\Psi(t)\right\rangle$, where $k$ is an integer, $T$ is the period of the driving, and $\varphi$ is the global phase, i.e., the state is frozen at multiple
periods. Therefore, when the CDT occurs, the vanishing of the quasi-energies can be directly determined by the constant population on every measurement basis after multiple periods. For this experiment, we first detected the $\left|+\right\rangle $ basis to identify the frozen point of $E$. Subsequently, we confirmed that the CDT indeed occurs by detecting the state evolution onto bases $\left|0\right\rangle $ and $\left|i\right\rangle $.

\bigskip
\noindent \textbf{Funding:} This work was supported by the National Key Research and Development Program of China (Nos. 2017YFA0304100, 2016YFA0302700), the National Natural Science Foundation of China (Nos.11874343, 61327901, 11774335, 11474270, 11734015, 11821404), Key Research Program of Frontier Sciences, CAS (No. QYZDY-SSW-SLH003), the Fundamental Research Funds for the Central Universities (Nos. WK2470000026,WK2470000027, WK2470000028), Anhui Initiative in Quantum Information Technologies (AHY020100, AHY070000).
\noindent \textbf{Author contributions:} YJH proposed this project, CFL and GCG conducted experiments; RH, MZA, JMC and YFH designed the experiment and collected the data, using an apparatus with significant contributions from RH, MZA and JMC; RH, MZA, YJH, JMC, TT and YFH prepared and wrote the manuscript. All authors have read and approved the final manuscript.
\noindent \textbf{Competing financial interests:} The authors declare no competing interests. \noindent \textbf{Data and materials availability:} All data needed to evaluate the
conclusions in the paper are present in the paper. Additional data related to this paper may be requested from the authors.

\end{document}